\documentclass[journal]{IEEEtran}
\usepackage{cite}
\usepackage{amsmath,amssymb,amsfonts}
\usepackage{hyperref}
\usepackage{graphicx}
\usepackage{textcomp}
\usepackage{bm}
\usepackage{relsize}
\usepackage{amsmath}
\usepackage{amsmath, amsthm, amssymb}
\usepackage{float}
\usepackage{graphicx}
\usepackage{epstopdf}
\usepackage{lipsum}
\usepackage{caption}
\usepackage{algorithm}
\usepackage{algpseudocode}
\usepackage{verbatim}
\usepackage{calligra}
\usepackage[font={footnotesize}]{caption}
\usepackage{subfigure}
\usepackage{xcolor}
\usepackage{mathtools}
\usepackage{stackengine}
\usepackage[scr=boondoxo,
scrscaled=1.15,]{mathalfa}
\usepackage{bbm}
\usepackage{adforn}
\usepackage{acronym} \input{Acronym}
\usepackage{tikz}
\definecolor{bg}{RGB}{237,239,223}
\definecolor{frame}{RGB}{104, 85, 95}

\DeclareMathAlphabet{\mathpzc}{OT1}{pzc}{m}{it}
\DeclareMathAlphabet{\mathcalligra}{T1}{calligra}{m}{n}

\makeatletter 
\newcommand\semiHuge{\@setfontsize\semiHuge{22.6}{27}}
\newcommand\semihuge{\@setfontsize\semihuge{18.93}{23.45}}

\begin{document}
%Dual-Functional Codebook Design for NF Integrated Sensing and Communication
%\title{Dual-Functional Codebook Design: Enabling Near-Field Sensing and Communication Synergy}

%\title{Near-Field ISAC: Synergy of Dual-Purpose Codebooks and Space-Time Adaptive Processing}

%\title{Holographic ISAC: Synergy of Near-Field Codebooks and  Space-Time Adaptive Processing}

\title{Near-Field ISAC: Synergy of Dual-Purpose Codebooks and Space-Time Adaptive Processing}

\author{
Ahmed Hussain, Asmaa Abdallah, Abdulkadir Celik, Ahmed M. Eltawil
%\vspace{-0.75cm}
}
\maketitle
%TC:ignore
\begin{abstract} 
Integrated sensing and communication (ISAC) has emerged as a transformative paradigm, enabling situationally aware and perceptive next-generation wireless networks through the co-design of shared network resources. With the adoption of millimeter-wave (mmWave) and terahertz (THz) frequency bands, ultra-massive MIMO (UM-MIMO) systems and holographic surfaces unlock the potential of near-field (NF) propagation, characterized by spherical wavefronts that facilitate beam manipulation in both angular and range domains. This paper presents a unified approach to near-field beam-training and sensing, introducing a dual-purpose codebook design that employs discrete Fourier transform (DFT)-based codebooks for coarse estimation of sensing parameters and polar codebooks for parameter refinement. Leveraging these range and angle estimates, a customized low-complexity space-time adaptive processing (STAP) technique is proposed for NF-ISAC to detect slow-moving targets and efficiently mitigate clutter. The interplay between codebooks and NF-STAP framework offers three key advantages: reduced communication beam training overhead, improved estimation accuracy, and minimal STAP computational complexity. Simulation results show that the proposed framework can reduce STAP complexity by three orders of magnitude, validating efficacy, and highlighting the potential of the proposed approach to seamlessly integrate NF communication and sensing functionalities in future wireless networks.
\end{abstract}
\IEEEpeerreviewmaketitle
%TC:endignore
 
\section*{\textbf{Introduction}}

\begin{figure*}
    \centering
    \includegraphics[width= 1.0 \textwidth]{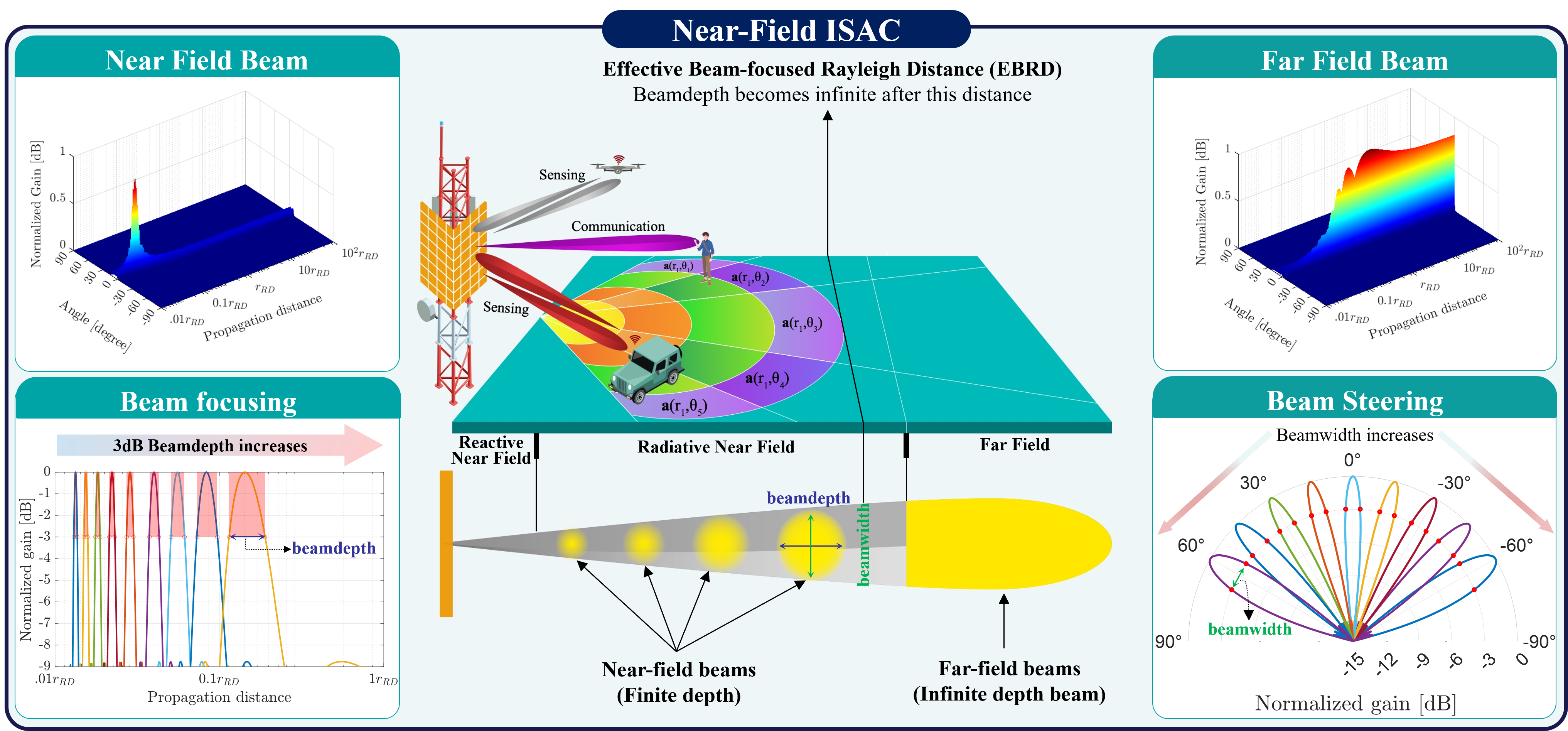}
    %\caption{Finite-depth beams characterize the NF, where the beam-depth increases with distance (leftmost diagram). The center diagram shows the electromagnetic field regions around a holographic antenna array, highlighting the transition between NF and FF zones. The rightmost diagram illustrates FF beams with infinite depth and beam broadening in the angle domain.}
    %TC:ignore
    \caption{Illustration of NF-ISAC concept and comparison between FF and NF beam characteristics.}
%TC:endignore
    \label{fig: Main_figure}
\end{figure*}

\IEEEPARstart{I}{ntegrated} sensing and communication (ISAC) systems are poised to be a cornerstone of future 6G networks, leveraging the abundant spectrum available in the millimeter-wave (mmWave) and terahertz (THz) bands to achieve unprecedented capacity enhancements \cite{Jiang2021Survey}. The shorter wavelengths at these frequencies facilitate the deployment of densely packed massive multi-input multi-output (mMIMO) systems that employ narrow, and highly directional beams to enhance spectral efficiency \cite{rappaport2019wireless}. However, as the beamwidth narrows, beam management becomes complex, requiring precise and robust beam training. Unlike sub-6 GHz systems, which feature wider beams and stochastic channel behaviors, mmWave/THz channels exhibit a more geometric and deterministic nature, offering superior sensing capabilities that can be exploited for enhanced beam management.

Building on these foundations, ultra mMIMO (UM-MIMO) systems push the boundaries of mMIMO by leveraging the unique properties of the radiative near-field (NF) that is upper-bounded by Rayleigh distance. Conventional mMIMO systems are primarily designed to operate beyond Rayleigh distance in the far-field (FF) region, wherein planar wavefronts allow beam management solely in the angular domain. Conversely, UM-MIMO's operation within the NF region is characterized by spherical wavefronts and requires beam training in both range and angle domains.

Traditional beam-training methods, such as those using discrete Fourier transform (DFT)-based codebooks in 5G networks, are designed for FF operations and fail to account for the spherical wavefronts within the NF region. Although polar codebooks have been proposed for NF communications to address this limitation, they often suffer from substantial overhead due to the need for beam-training in both angular and distance domains. {To reduce the complexity of polar codebook-based methods, \cite{9913211} proposes a two-stage approach using DFT for angle estimation only and a polar codebook for range refinement. Hierarchical designs like \cite{10365224} lower overhead via multi-resolution codebooks but require iterative feedback. Subarray-based methods \cite{10239282} estimate coarse direction under FF assumptions and refine with polar codebooks, though reduced subarray size leads to wider beams and degraded spatial resolution.} At this intersection, the co-design of NF and ISAC systems offers a promising path toward reducing overhead while maintaining high accuracy and efficiency for both communication and sensing tasks. For instance, 5G reference signals, e.g., the synchronization signal block (SSB), demonstrated sensing potential thanks to their periodic beam-sweeping characteristics \cite{golzadeh2023downlink}. Minimal modifications to standards are needed to dual-purpose design and use of such control signals, highlighting the broader impact of seamless integration and efficient resource utilization.

Inspired by these points, this paper introduces a unified approach for NF beam-training and sensing operations. We propose a hybrid approach wherein a DFT-based codebook is initially employed to identify potential users and obtain coarse range and angle estimates over a wide coverage area. These initial estimates are then refined using a polar codebook, effectively reducing training overhead while enhancing accuracy. Leveraging the refined estimates from the polar domain, a space-time adaptive processing (STAP) technique customized for NF-ISAC is devised to detect slow-moving targets while effectively mitigating clutter by jointly exploiting spatial and temporal dimensions. Therefore, developed STAP approach not only provides accurate estimate of sensing parameters (i.e., range, angle, Doppler), but also minimizes the beam-training overhead as beam-focusing can be swiftly and precisely adjusted to the estimated range and angle of the users. Since STAP is computationally intensive, the refined estimates derived from the polar codebooks are particularly valuable in narrowing the search space, thereby reducing overall computational complexity by three orders of magnitude and enhancing system efficiency.

\section*{\textbf{Preliminaries of NF Communication}}
This section first delves into the electromagnetic (EM) properties of the NF region and its unique resolving capabilities compared to the FF region, then presents a primer on DFT and polar codebooks.
%Subsequently, we introduce DFT and polar codebooks and highlight their roles in channel estimation through beam-training, which is crucial for effective NF-ISAC.

\subsection*{\hspace{7 pt} \textbf{A Comparative Analysis of Beam Pattern Characteristics}}

\subsubsection*{\textbf{FF Beam-Steering vs. NF Beam-Focusing}}
Unlike conventional FF beam-steering in angular domain, NF beam-focusing exploits the spherical wavefronts of NF propagation to manipulate both beam angle and range. The $3$dB beamdepth is defined as the distance interval along the radial direction where the beam maintains at least half of its maximum gain. As illustrated in Fig. \ref{fig: Main_figure}, finite beamdepth enables the generation of multiple beams at varying ranges within the same angular direction \cite{10443535}, thereby enhancing spatial multiplexing gains. Furthermore, this additional degree of freedom allows the resolution of users in the distance domain, thereby advancing both localization accuracy and communication performance.

% In the NF, finite beam-depth enables the generation of multiple beams at varying ranges within the same angular direction, as illustrated in Fig. \ref{fig: Main_figure}. This capability, known as beam-focusing, enhances spatial multiplexing by allowing simultaneous communication with multiple users positioned at different distances along the same direction \cite{kosasih2024finite}. Unlike traditional beam steering, which focuses solely on angular directionality, beam-focusing leverages the spherical wavefronts of NF propagation to include range information. This additional dimension can be exploited to accurately estimate the distance of users, providing enhanced resolution for localization and communication tasks.

\subsubsection*{\textbf{Beam Broadening}}
While the FF beamwidth expands as the beam is steered away from the array boresight [bottom-right, Fig. \ref{fig: Main_figure}], the NF beam-depth enlarges as the focal region moves farther from the array, increasing quadratically with the focus distance [bottom-left, Fig.\ref{fig: Main_figure}]. Furthermore, beam-depth is smallest at the array boresight and becomes progressively larger toward the off-boresight directions \cite{hussain2024near,10934779}, as the effective aperture of the array reduces at large angles leading to reduced gains.

%In the FF, beam-depth increases as the beam is steered away from the array boresight as depicted in bottom-right diagram of Fig. \ref{fig: Main_figure}. Similarly, the beam-depth enlarges (bottom-left diagram of Fig.\ref{fig: Main_figure}) as the focal region moves farther from the array, increasing quadratically with the focus distance. Furthermore, beam-depth is smallest at the array boresight and becomes progressively larger toward the off-boresight directions \cite{hussain2024near,asmaa2024near}.

\subsubsection*{\textbf{Lateral vs. Axial Resolution}}
 Lateral resolution, a.k.a., cross-range resolution, refers to the ability of \textit{distinguishing two targets at the same range but different angles}. It depends on both the beamwidth and the distance from the antenna array \cite{627147, hussain2024near}. Although beamwidth is constant in both NF and FF, the proximity of the NF improves lateral resolution. Moreover, NF beams' finite beam-depth introduces axial resolution, enabling \textit{discrimination between two targets aligned along the same angle}, provided their separation exceeds the beam-depth. This dual-resolution capability highlights the unique advantages of NF beamforming for ISAC. 

 \subsubsection*{\textbf{Effective NF Boundaries}}
Beam-focusing is achieved only when the focal point lies within a boundary defined by the effective beam-focused Rayleigh distance (EBRD), which depends on the frequency, aperture size, spatial focal region, and array geometry \cite{ahmed2024near}. For a given array geometry, the EBRD increases with the aperture size, which can be expanded by increasing the number of antenna elements or the inter-element spacing. Interestingly, the EBRD is smallest for a square array and reaches its maximum when the uniform rectangular array approaches a uniform linear array (ULA) configuration. 
%For a ULA, the EBRD can be calculated as $\frac{r_{\mathrm{RD}} \cos^2\theta}{7}$, where $r_{\mathrm{RD}}$ denotes the Rayleigh distance.

\subsection*{\hspace{7 pt} \textbf{FF Codebooks vs. NF Codebooks}}
% Beam management, particularly for UM-MIMO systems operating at mmWave frequencies, is an efficient approach to acquire CSI. While traditional 5G systems rely on DFT codebooks, emerging research indicates that polar codebooks may be more suitable for NF beam management.

\subsubsection*{\textbf{DFT Codebooks}}
{5G beam training typically involves sweeping a grid of DFT beams, with users reporting the index corresponding to the beam with the highest received gain. To counteract misalignment caused by mobility, blockage, or device rotation, beam tracking and refinement are performed using narrower, more focused beams. The DFT codebook remains a practical choice in 5G systems due to its FFT-compatible structure, low implementation complexity, and suitability for hybrid beamforming. While standard DFT beams provide coarse localization, oversampled variants support fine-grained beam tracking.}

\subsubsection*{\textbf{Polar Codebooks}}
{NF beam alignment poses challenges due to the need for sweeping beams across both angular and distance domains, leading to high complexity and training overhead. 
%Polar codebook designs have recently been explored to mitigate these issues. 
Polar codebooks are designed based on the Rayleigh distance, which often overestimates the NF region in terms of beam-focusing and multiplexing gains \cite{hussain2024near,ahmed2024near}. This overestimation leads to excessive sampling and unnecessarily large codebook sizes.
%Alternatively, NF codewords spaced according to the beam-depth and sampled within the EBRD region exhibit low spatial correlation \cite{ahmed2024near}. These optimized polar codebooks provide superior performance while significantly reducing computational overhead.
As a remedy, recent research has explored NF codebooks that are optimized based on beam-depth and sampled within the EBRD region \cite{ahmed2024near}, yielding lower spatial correlation between codewords and superior performance while significantly reducing beam-training overhead.} 

% Nevertheless, the NF regime inherently incurs significant training overhead due to embedded range information. 
%Nevertheless, fundamental model parameters that are range-independent in the FF, such as effective aperture area, array gain and polarization become location-dependent in the NF, making communication system performance highly dependent on the range information. To address these challenges, NF communication systems must leverage advanced sensing paradigms to adapt to the dynamic electromagnetic environment and reduce training overhead.

It is worth noting that fundamental system parameters that are range-independent in the FF (e.g., effective aperture area, array gain, and polarization) become highly location-dependent in the NF. This dependency makes communication system performance sensitive to accurate range estimation. Mitigating these challenges necessitates the symbiosis of NF communication and sensing, which is addressed next.

\begin{figure*}[t!]
    \centering
    \includegraphics[width=\textwidth] {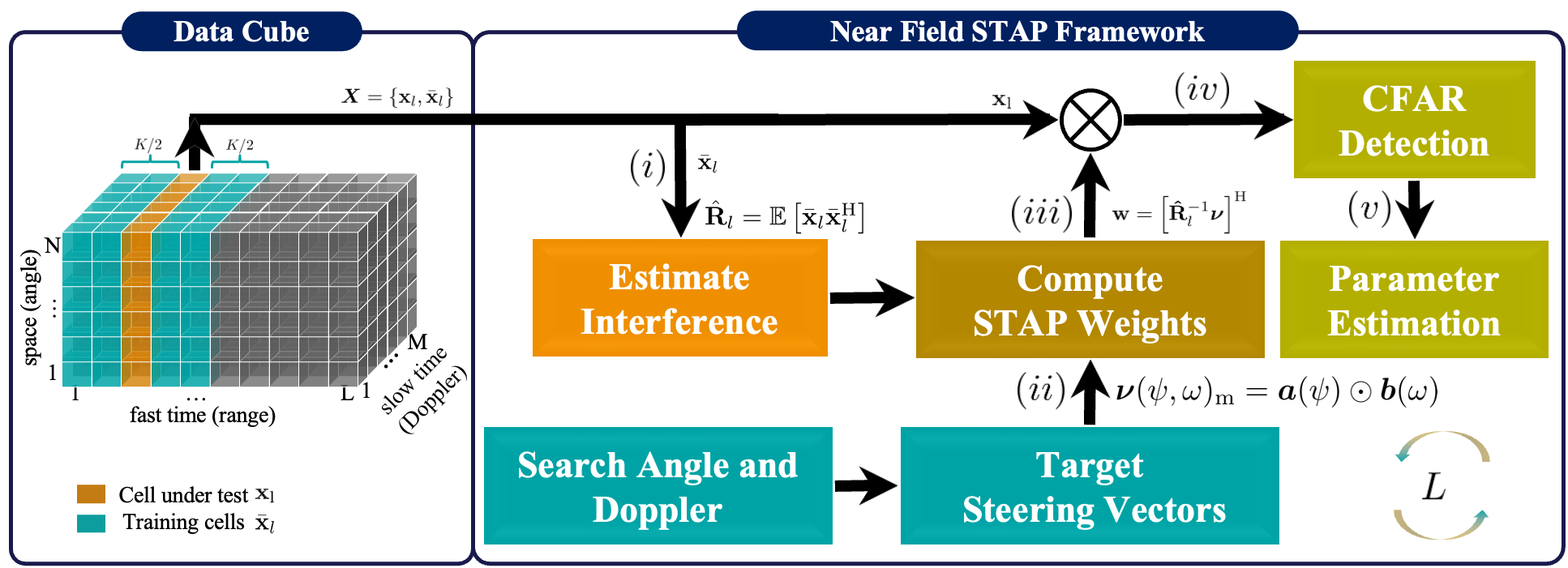}
%TC:ignore
    \caption{Illustration of radar data cube and flow diagram of the proposed NF-STAP algorithm.} %TC:endignore
    \label{fig:STAP_framework}
\end{figure*}

%%%%%%%%%%%%%%%%%%%%%%%%%%%%%%%%%%%%%%%%%%%%%%%%%%%%%
%%%%%%<----------SECTION III ----------->%%%%%%
%%%%%%%%%%%%%%%%%%%%%%%%%%%%%%%%%%%%%%%%%%%%%%%%%%%%%
\section*{\textbf{Space-Time Adaptive Processing towards Sensing-aided NF Communication}}

Future NF-ISAC systems must be capable of operating effectively in cluttered and interference-prone environments, which pose daunting challenges often overlooked in current ISAC research. Ground clutter, particularly extended in both angle and range dimensions, hinders accurate target detection. In radar look-down modes, ground clutter often dominates Gaussian noise as the primary source of interference. The situation is further complicated when detecting low-velocity targets, e.g., pedestrians or cyclists, due to their Doppler signatures being close to the mainlobe clutter. Conventional radar signal processing algorithms lack the versatility to mitigate such interference effectively, emphasizing the need for adaptive techniques such as STAP. In the following, we propose a NF-STAP framework designed to address these challenges, demonstrating how it can seamlessly integrate with and enhance critical NF communication performance metrics.

\subsection*{\hspace{7pt} \textbf{STAP Overview}}

The most fundamental part of STAP algorithm is the generation of radar data cube. Let us consider an $N$-element monostatic radar, transmitting $M$ pulses at a pulse repetition frequency (PRF) of $f_r$. The total duration of the $M$ pulses $\frac{M}{f_r}$ is referred to as the coherent processing interval (CPI). During this interval, the received signals are down-converted, and stored as digital IQ samples in a three-dimensional radar data cube as shown in Fig. \ref{fig:STAP_framework} and explained below:
    \begin{itemize}
        \item \textbf{Fast time} represents range bins corresponding to the sampled received signals. The number of range bins equal, $L = f_s/f_r$, where $f_s$ is the sampling rate and $f_r$ represents the symbol rate in case of OFDM radar.
        \item \textbf{Slow time} corresponds to the $M$ pulses within a CPI. The desired Doppler resolution, $\frac{f_r}{M}$, dictates the number of pulses or symbols.
        \item \textbf{Antenna number} 
        %represent the spatial dimension of the data cube which holds the data samples with respect to different antenna elements. 
        represents the spatial dimension, storing samples collected from $N$ different antenna elements.
    \end{itemize}

%The STAP framework, depicted in Fig. \ref{fig:STAP_framework}, integrates spatial and temporal signal processing to detect low-velocity targets obscured by clutter. By leveraging data collected from an antenna array over multiple pulses during a CPI, STAP performs two-dimensional adaptive filtering to enhance radar performance significantly \cite{khan2020adaptive}.
As illustrated in Fig. \ref{fig:STAP_framework}, STAP algorithm leverages the data cube in the following steps, which is repeated for each range bin, $l \in [1,L]$, \cite{khan2020adaptive}: (i) Clutter and interference are characterized through a covariance matrix, $\hat{\mathbf{R}}_l$, which is estimated for the cell under test $\mathbf{x}_l \in \mathbb{C}^{MN \times 1}$ by averaging interference from $K$ training cells. Thus, it is defined by $\hat{\mathbf{R}}_l=\mathbb{E}[\bar{\mathbf{x}}_l \bar{\mathbf{x}}_l^\mathrm{H}]$, wherein $\bar{\mathbf{x}}_{l} \in \mathbb{C}^{MN \times 1}$, $\forall \ l \in [l-K/2,\cdots,l-1,l+1,\cdots,l+K/2]$; (ii) The STAP process evaluates the range bin $\mathbf{x}_l$ across all potential Doppler frequencies $\omega$ and look angles $\psi$, represented by the target steering vector $\boldsymbol{\nu}(\psi,\omega)$; (iii) For each Doppler and angle value, the framework dynamically computes weight vectors $\mathbf{w} = [\hat{\mathbf{R}}_l^{-1}\boldsymbol{\nu}]^\mathrm{H}$, optimized to maximize the signal-to-interference-plus-noise ratio (SINR); (iv) The weight vector is applied to $\mathbf{x}_l$ as $\mathbf{w}\mathbf{x}_l$ to filter out interference while maximizing the target signal, yielding a scalar value for a given Doppler and angle pair; and (v) The STAP outputs for all Doppler and angle values are then passed to a {constant false alarm rate (CFAR)} detector, which determines the presence or absence of a target based on a predefined false alarm probability. Following the CFAR detection, angle and Doppler parameters are estimated, and the data is further processed for target tracking and localization. Since steps (i)-(v) are repeated for all $L$ range bins, STAP incurs an overall time-complexity of $\mathcal{O}\left( L (MN)^3 \right)$.

\subsection*{\hspace{7pt} \textbf{How to Adopt STAP for NF Sensing}} 
The planar wavefront assumption underlying the FF target steering vector does not hold in the NF. In the FF, the target steering vector is constructed as the Kronecker product of the FF spatial and the Doppler steering vector, where the latter captures phase transitions caused by target motion between pulses. In contrast, as depicted in Fig. \ref{fig:STAP_framework}, the NF spatial steering vector $\mathbf{a}(\psi)$ is based on the spherical wavefront, accounting for the varying distances $r^{(n)}$ between each antenna and the focal point. Additionally, the NF effect causes Doppler frequency to vary across the antenna aperture. This variation is captured by the normalized non-uniform Doppler, $\frac{v^{(n)}}{f_r}$, where $v^{(n)}$ represents the velocity observed by the $n^{\mathrm{th}}$ antenna element. The NF target steering vector ${\boldsymbol{\nu}}$ is thus computed as the Hadamard product of $\mathbf{b}(\omega)$ and $\mathbf{a}(\psi)$, calculated over $M$ pulses within a CPI. The subsequent steps of weight vector calculation, CFAR processing and parameter estimation are similar to the FF case. 

%\subsubsection*{\hspace{7pt} \textbf{Challenges}} 
It is worth noting that the actual interference covariance matrix $\mathbf{R}$, is often unavailable and must be estimated from surrounding training cells. The accuracy of this estimation depends on the availability of homogeneous training cells. However, in practical scenarios, this assumption may not always hold due to multiple reflections, and the presence of secondary targets within the training cells. FF covariance matrices exhibit a convenient Toeplitz-block-Toeplitz structure, which is disrupted by spherical wavefronts in the NF. Moreover, the substantial dimensions of the covariance matrix $\mathbf{R}$, with a size of $(NM \times NM)$, render direct covariance estimation computationally intractable. Consequently, the time complexity of NF-STAP, given by $\mathcal{O}\left( L (MN)^3 \right)$, becomes prohibitively high, particularly in the context of UM-MIMO, where $N$ is very large.

% Furthermore, the large dimensions of the covariance matrix $\mathbf{R}$, which is of size $(NM \times NM)$ ,render covariance estimation computationally prohibitive. Accordingly, the time complexity $\mathcal{O}\left( L (MN)^3 \right)$ exacerbates in NF-STAP, with $N$ being very large for UM-MIMO.

\subsection*{\hspace{7pt} \textbf{Sensing-aided NF Communication}} \label{sec:Convergence_of_Sensing_and_Communication}
At the cost of additional complexities, NF-STAP offers two distinct advantages over its FF counterpart: 

\subsubsection*{ {\large \adforn{3}}  \textbf{Interference Suppression and Ambiguity Reduction}} 
In traditional FF radar systems, interference often spreads across multiple range bins, complicating clutter and jammer mitigation. %However, in NF radar, the beam-focusing property enables precise spatial targeting, allowing interference from clutter and jammers to be proactively suppressed. The proximity of NF coupled with beam-focusing property also mitigates range and Doppler ambiguities, as the radar predominantly receives echoes from the focused location.
NF radar, however, leverages its beam-focusing capability to achieve precise spatial targeting and minimize detrimental impacts of clutter and jammers. Moreover, the proximity inherent to NF operation mitigates range and Doppler ambiguities, as echoes are predominantly received from the focal region. All these characteristics enhance interference suppression while improving identification and detection accuracy at the same time. 

\subsubsection*{{\large \adforn{3}} \textbf{Accurate Range Estimation}}
Axial range resolution in the NF is constrained by the beam-depth and becomes less precise at greater distances from the base station (BS). To enhance range resolution, multi-carrier OFDM radar can be employed, achieving a resolution of $\frac{c}{2B}$, where $B$ is the bandwidth. For a specific angle, a broad beam in the range dimension can be designed to illuminate the NF. 

%While broad beam techniques—using dual polarization, amplitude weighting, phase tapering, and antenna selection—are well-established for FF systems, their application in the NF remains largely unexplored. Generating a single broad beam can maximize the SNR in the NF and reduce the required transmission resources for Doppler estimation. 

The better interference mitigation, ambiguity reduction, and range estimation provided by NF sensing eventually paves the way for a boost in NF communication performance as follows:

\subsubsection*{$\checkmark$  \textbf{Blockage Prediction and Proactive Beamforming}}
Mobility scenarios at mmWave/THz networks introduce substantial Doppler effects and necessitate frequent feedback updates during beam tracking, thereby degrading communication system performance. %Additionally, mmWave frequencies, operating primarily in line-of-sight (LoS) conditions, are vulnerable to minor blockages.
Also considering their vulnerability to minor blockages, sensing-aided NF communication can proactively align beams by predicting blockages to reduce feedback overhead and improve robustness.
%Sensing-assisted predictive beamforming, enabled by the ISAC framework, mitigates these challenges by proactively aligning beams and reducing feedback overhead. 
Unlike FF systems' limitation to radial velocity estimation, an additional NF sensing advantage lies in its ability to estimate both radial and transverse velocity components of moving targets. By exploiting the non-uniform Doppler frequencies arising from spherical wavefronts, NF systems can more accurately track target motion and predict their trajectories to enhance the reliability of mmWave/THz links in dynamic environments. 

 \subsubsection*{ $\checkmark$ \textbf{Site-Specific Beamforming and Codebook Design}}
NF sensing can be employed to generate channel knowledge maps, identifying potential obstructions within the propagation path. Furthermore, these maps provide critical information about the propagation environment, including the location of obstacles, scatterers, dominant reflectors, and potential line-of-sight and non-line-of-sight paths. By analyzing this environmental data, the system can predefine high-probability beam directions and focus the beam alignment process on areas most likely to yield high signal quality. This approach avoids the exhaustive search over all possible beam directions, significantly reducing the time and computational overhead required for initial access.

%This information can be utilized to reduce the excessive beam sweeps required for initial access. 

\begin{figure*}[t!]
    \centering
    \includegraphics[width= \textwidth] {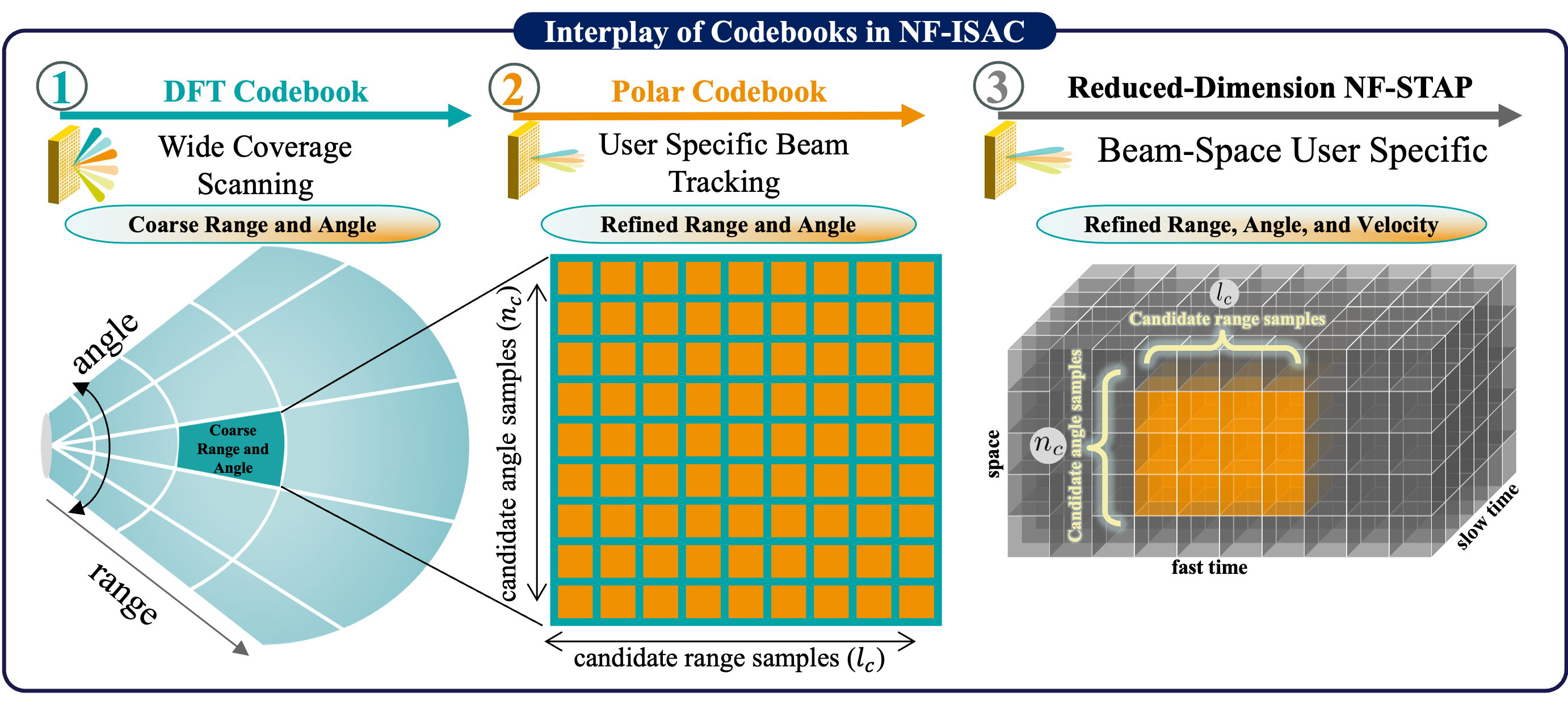}
%TC:ignore
     \caption{Demonstration of NF-Sensing complexity reduction through the interplay of DFT codebooks, polar codebooks, and NF-STAP.}
%TC:endignore
    \label{fig:interplay_Codebooks}
\end{figure*}

%%%%%%%%%%%%%%%%%%%%%%%%%%%%%%%%%%%%%%%%%%%%%%%%%%%%%
%%%%%%<----------SECTION IV ----------->%%%%%%
%%%%%%%%%%%%%%%%%%%%%%%%%%%%%%%%%%%%%%%%%%%%%%%%%%%%%

\section*{\textbf{Communication-aided NF Sensing: \\ Interplay of Codebooks and NF-STAP}}

A cross-field beam-training solution that accommodates both FF and NF users is highly desirable for future networks. However, the training overhead remains a significant challenge, particularly for NF users. To this aim, we propose a unified framework that capitalizes on the interplay between communication codebook training and NF-STAP, thereby unleashing the full potential of NF communication and sensing synergy. As illustrated in Fig. \ref{fig:interplay_Codebooks}, the developed framework is executed through the following three steps:

%By capitalizing on the distinctive capabilities of NF-assisted sensing and sensing-assisted communication, as elaborated in the preceding section, the training overhead can be alleviated and the computational burden of STAP can also be reduced. %By integrating sensing and communication paradigms, it is possible to reduce beam sweeping resources while improving estimation performance, as illustrated in Fig. \ref{fig:interplay_Codebooks}. The NF ISAC-based beam-training solution comprises three phases:

\subsubsection*{\hspace{-10pt}{\textcircled{\footnotesize \textbf{1}}} \textbf{UE Discovery \& Coarse Sensing using DFT Codebooks}}
%DFT codebooks are employed for initial wide-area scanning, enabling coarse estimation of range and angle parameters for both NF and FF users. This approach offers the following advantages:
In 5G networks, DFT codebooks are periodically employed for wide-area sweeping to facilitate initial access and synchronization, which can also be used to coarsely estimate UEs' range and angle parameters \cite{hussain2025near}, offering two key advantages for NF communications:
\begin{itemize}
    \item[{\large \adforn{3}}] Its search is limited to the angular domain and avoids the computational overhead of a two-dimensional angle-range search in NF scenarios.
    \item[{\large \adforn{3}}] It can simplify the initial access process for NF users by providing coarse range and angle estimates. 
\end{itemize}

\begin{figure*}[t!]
    \centering
    \includegraphics[width= \textwidth] {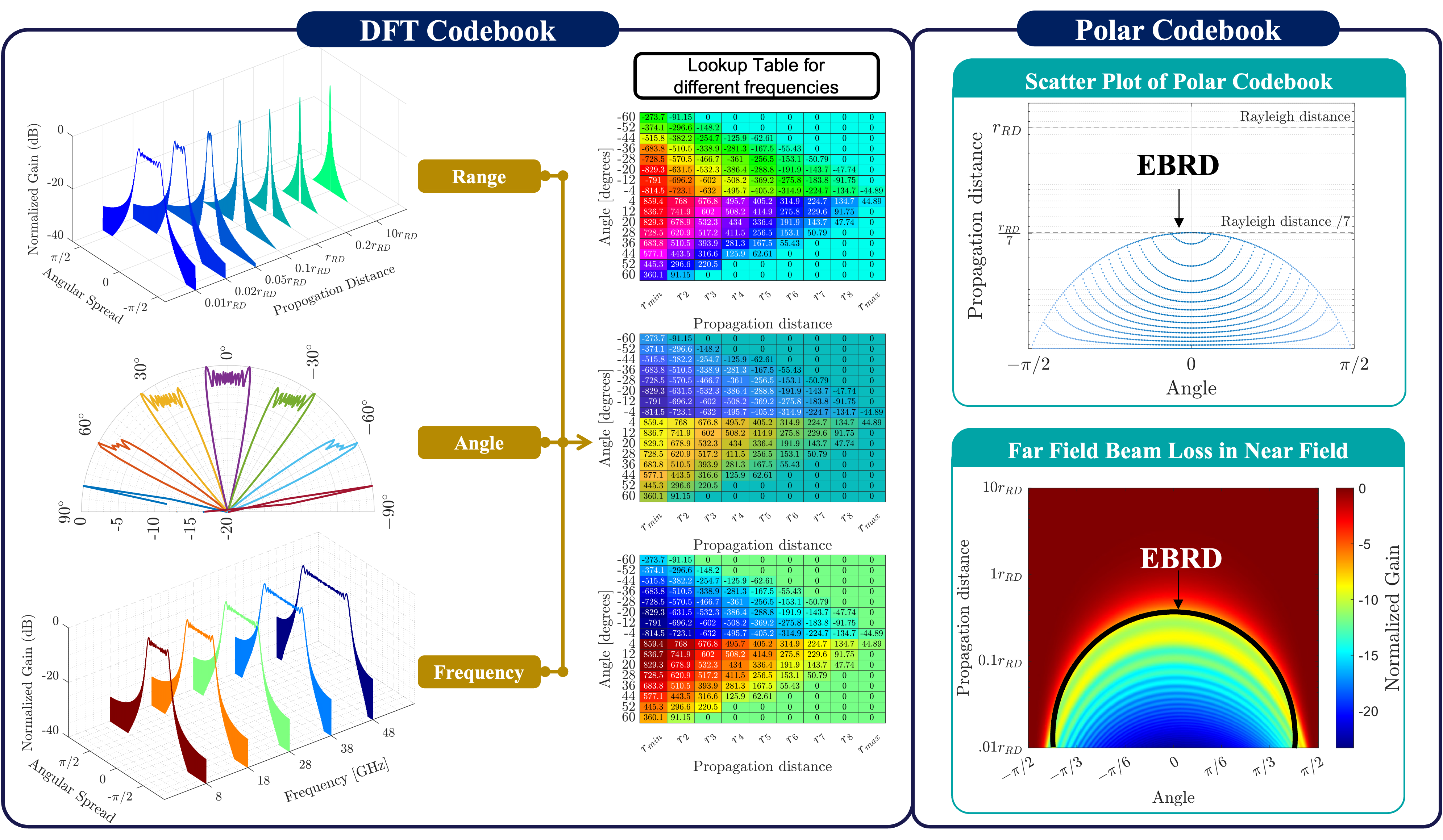}
 %TC:ignore
    \caption{Illustration of DFT angular spread in the NF region and polar codebooks for an ULA operating at $28$ GHz with 256 elements.}
    %TC:endignore
    \label{fig:Comm_Codebooks}
\end{figure*}

DFT beams are transmitted uniformly across the spatial domain to provide sufficient gain for FF users, which typically experience peak gain from a single DFT beam and report the corresponding beam index to the BS.
%For FF users, a single DFT beam typically provides sufficient gain, allowing them to report their beam index to the base station (BS). 
NF users, however, experience gain over multiple DFT beams due to spherical wavefront characteristics, which are approximated as a superposition of planar wavefronts. Since the wavefront curvature varies with both distance and angle relative to the BS, it causes the signal to span a broader range of angles as it propagates, resulting in \textit{angular spread} as shown in Fig. \ref{fig:Comm_Codebooks}, where the angular spread is uniquely mapped to UE position for a given carrier frequency \cite{hussain2025near}. It increases as the user moves closer to the BS, reflecting the pronounced NF effects. Additionally, higher carrier frequencies amplify these effects, further widening the angular spread. For a fixed distance, the angular spread is narrower in the endfire directions compared to the broadside directions.

The range-, angle-, and frequency-dependent characteristics of angular spread can be leveraged to estimate the position of user. To this aim, a lookup table can be precomputed to encapsulate the angular spread for each polar point at a given frequency, as illustrated in Fig. \ref{fig:Comm_Codebooks} (leftmost). The lookup table is constructed by first defining a polar grid and then computing the angular spread at each grid point. In this grid, angles are sampled uniformly, while range points are selected based on beamdepth. During DFT beam sweeping, the UEs measure the angular spread and match it with the entries in their lookup table to estimate both range and angle. The UE then reports these parameters to the BS for further processing. Although this method enables efficient initial access, the NF beam-depth limits the range resolution provided by the DFT-based angular spread estimation. Additionally, the broad angular beams used for initial access may lead to overall coarse range and angle estimates.

 \subsubsection*{ \hspace{-14pt}{\textcircled{\footnotesize \textbf{2}}} \textbf{Refined Sensing \& Beam Tracking with Polar Codebook}}
After users are identified through the DFT codebook, polar codebooks can be deployed for user-specific beam-training and tracking operations. FF beams experience significant attenuation, often exceeding 3 dB, in the NF effective region as depicted in Fig. \ref{fig:Comm_Codebooks} (bottom-right), making polar codebooks essential for precise beam alignment and reliable data transmission. Fig. \ref{fig:Comm_Codebooks} (top-right) illustrates a scatter plot of a polar codebook for a ULA, where samples in the distance domain are separated by the beam-depth and limited to the EBRD boundary. These polar codebooks refine the coarse range and angle estimates obtained during DFT-based training by oversampling beams within the candidate polar region. The polar codebooks provide $l_c$ candidate range samples centered around the previously obtained coarse range sample, and $n_c$ candidate angle samples centered around the previously obtained coarse angle sample. To reduce the impact of initial estimation errors, the parameters $l_c$ and $n_c$ can be increased. This refinement ensures the generation of highly directional beams tailored to each UE. Beam tracking is achieved by probing neighboring polar regions to compensate for variations in the optimal transmission direction due to user movement or environmental changes. For NF beam tracking in high-mobility scenarios, Doppler estimates derived from NF-STAP can be utilized for predictive beamforming, mitigating overhead.

\subsubsection*{ \hspace{-20 pt} {\textcircled{\footnotesize \textbf{3}}} \textbf{Low-Complexity NF-STAP}}
 The final step provides refined range and Doppler estimates, while the initial parameter estimates from the polar codebook reduce the computational complexity of NF-STAP by narrowing the search space. Specifically, the complexity is reduced from $\mathcal{O}\left( L (MN)^3 \right)$ to $\mathcal{O}\left( l_c (Mn_c)^3 \right)$. %, where $l_c$ and $n_c$ represent the number of candidate range and angle samples, respectively. 
 Quantitatively, the reduction factor is surely $\frac{LN^3}{l_c n_c^3} \geq 1000$, given that $N$ and $L$ are on the order of hundreds, while $l_c$ and $n_c$ are on the order of tens. This approach optimizes the beam-training process and enhances overall system performance. By judiciously combining DFT and polar codebooks, along with NF-assisted sensing and communication, this proposed beam-training solution effectively addresses the challenges of NF beam-training, reducing overhead and optimizing coordination gain of ISAC.

\begin{figure*}[t!]
    \centering
    \includegraphics[width= \textwidth] {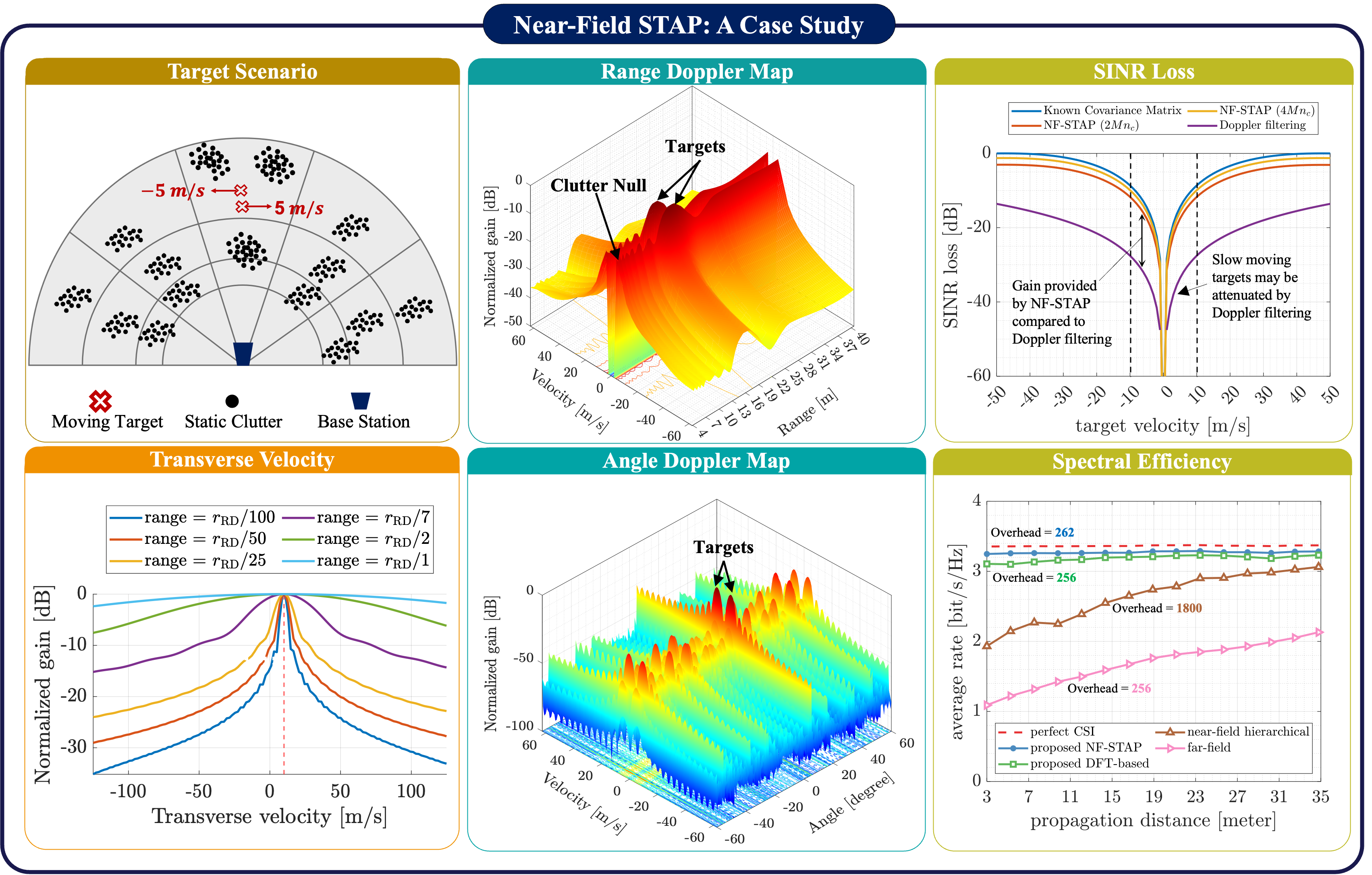}
    %TC:ignore
    \caption{The target illumination scenario is shown in the top-left diagram. After STAP processing, the range-Doppler and angle-Doppler maps are presented in the top-center and bottom-center diagrams, respectively. The bottom-left diagram illustrates the reduction in transverse velocity resolution as the distance increases within the near-field region. The top-right diagram shows the SINR loss as a function of target radial velocity, while the bottom-right diagram depicts the average rate versus propagation distance.}
    %TC:endignore
    \label{fig:NF_STAP_results}
\end{figure*}

\subsection*{\hspace{7pt} \textbf{NF-STAP Case Study}}
Reduced-dimension NF-STAP can mitigate the limitations of polar codebooks by providing refined range and Doppler estimates for moving targets. The estimated Doppler information can be leveraged for predictive beamforming, reducing the need for excessive feedback during beam tracking. To exemplify the NF-STAP, we consider an OFDM radar operating at 28 GHz and equipped with $N = 256$ antenna elements. During a CPI, the system transmits a total of $M = 128$ pulses / symbols. The target illumination scenario and associated results are illustrated in Fig. \ref{fig:NF_STAP_results}. Static scatterers, modeled as clutter, are present across all angle and range bins. Additionally, two targets, moving at velocities of $\pm 5 \, \text{m/s}$ in opposite directions, are positioned at the same angle of $5^\circ $. These targets are located at ranges of 15 m and 16 m, making them indistinguishable by the NF communication beam. 

After applying reduced-dimension NF-STAP, the range-Doppler map reveals a null around zero Doppler, highlighting the suppression of clutter. Furthermore, target peaks can be observed at approximately 15 m and 16 m. A bandwidth of $400 \, \text{MHz}$, providing a range resolution of $0.375 \ \text{m}$ , allows for the discrimination of the two targets. In the angle-Doppler map, the targets are observed at the same angle of $5^\circ$ but with opposite velocities of $5 \ \text{m/s}$ and $-5 \ \text{m/s}$.

As shown in top-right diagram of Fig. \ref{fig:NF_STAP_results}, the proposed NF-STAP achieves an SINR gain of $15$ dB compared to conventional Doppler filtering, enabling the detection of slow-moving targets. Furthermore, with $4Mn_c $ training samples where $n_c=8$, the proposed NF-STAP closely approaches the optimal STAP performance, assuming a known interference covariance matrix. 

The bottom-right diagram in Fig. \ref{fig:NF_STAP_results} depicts the average rate performance as a function of near-field communication distance. It can be observed that the proposed NF-STAP and DFT-based schemes outperform the near-field hierarchical \cite{10365224} and conventional far-field methods. Furthermore, the training overhead of the proposed DFT-based scheme matches that of the far-field scheme. By leveraging six additional beams, NF-STAP further refines the estimates, bringing its performance closer to that of perfect CSI.

It is worth noting that while transverse velocity can be resolved in the NF region—unlike in the FF—this capability is limited by the effective NF region, defined by EBRD. As shown in the bottom-left diagram of Fig. \ref{fig:NF_STAP_results}, the resolution of transverse velocity decreases as the distance between the BS and the targets increases. This behavior is expected since NF effects, such as angular spread, beam-focusing, and degrees of freedom, are confined to the specific portion of the NF region defined by the EBRD \cite{ahmed2024near}.

%It is worth noting that although the transverse velocity can be be resolved in the NF region unlike the case in FF region, The  ability to resolve transverse velocity in the NF is constrained by the effective NF region defined by the EBRD. As illustrated in the bottom-left diagram of Fig. \ref{fig:NF_STAP_results}, the resolution of transverse velocity diminishes with increasing distance between BS and the targets. This outcome is expected, as NF effects like angular spread, beam-focusing, and degrees of freedom are restricted to a specific portion of the NF region defined by the EBRD.

%%%%%%%%%%%%%%%%%%%%%%%%%%%%%%%%%%%%%%%%%%%%%%%%%%%%%
%%%%%%<----------SECTION VI: Conclusion and Future Directions ----------->%%%%%%
%%%%%%%%%%%%%%%%%%%%%%%%%%%%%%%%%%%%%%%%%%%%%%%%%%%%%

\section*{\textbf{Conclusion and Future Directions}}
This paper investigated codebook design for NF-ISAC, aiming to optimize the efficiency of NF beam training. The DFT codebook exploits angular spread to obtain coarse range-angle estimates with reduced overhead. Meanwhile, NF-STAP leverages candidate range-angle samples from the polar codebook to achieve high-resolution range and Doppler estimates while significantly reducing the computational complexity of traditional STAP operations. Future research directions are outlined in the sequel to advance NF-ISAC technologies:

\subsubsection*{{\hspace{-10 pt} {\large \adforn{72}} \textbf{Hardware Considerations}}} 
{In hybrid precoding systems, accurate phase and amplitude calibration may be incorporated during the construction of the DFT-based lookup table. Moreover, to account for a multitude of real-world impairments—such as synchronization errors, calibration inaccuracies, and mutual coupling—deep learning-based solutions may be required to replace the proposed ookup tables, as these effects can induce highly complex and non-ideal spatial signatures.} {Recently, rotatable and movable antenna architectures have emerged, offering flexible beamforming for precise beam alignment and dynamic interference suppression \cite{zheng2025rotatable}}. Moreover, to address the challenges posed by heterogeneous training cells in NF-STAP, techniques such as Doppler compensation, reduced-dimension STAP, diagonal loading, or leveraging structure in the training data can be employed \cite{khan2020adaptive}.

% \subsubsection*{{\hspace{-10 pt} {\large \adforn{72}} \textbf{Advanced Waveforms for High Mobility}}} 
% Emerging waveforms, such as orthogonal time-frequency space (OTFS) modulation, aim to overcome inter-carrier interference issues with traditional OFDM in high mobility environments \cite{rou2024orthogonal}. These waveforms not only mitigate the adverse effects of double selectivity but also leverage them for sensing and parameter estimation. Investigating the synergy between OTFS and NF phenomena presents exciting opportunities to enhance sensing capabilities and achieve superior performance in NF-ISAC systems. 

% In the NF, the Doppler dimension becomes coupled with the spatial dimension, fundamentally altering channel representations. Specifically, NF effects transform the angular-delay domain channel matrices, where different delays correspond to varying angular spreads, resulting in a strong angular-delay correlation. 
\subsubsection*{{\hspace{-10 pt} {\large \adforn{72}} \textbf{Holographic Surfaces}}} 
Holographic surfaces can dynamically manipulate EM waves at a fine-grained level. By applying the sampling principle, continuous holographic surfaces can be quantized with minimal inter-element spacing, allowing for more accurate modeling and enhanced control over beam depth and beam direction. However, this quantization significantly increases system complexity, as a vast number of discrete elements are required. Therefore, it is essential to leverage NF-ISAC functionalities to design site-specific codebooks tailored to holographic surfaces, thereby reducing codebook size and training overhead while incorporating intricate EM effects, such as leakage currents and mutual coupling. The proposed interplay between codebooks and NF-STAP can further mitigate the extensive beam-training overhead by efficiently directing signals to desired locations and leveraging the superposition of holographic patterns while accounting for leaky power constraints.

\subsubsection*{{\hspace{-10 pt} {\large \adforn{72}} \textbf{Integration of Mutli-modal Sensory Data}}} {By fusing diverse sensory inputs as in \cite{MF2024trans}, such as camera images, LiDAR point clouds, and position information, the beam-training overhead can be significantly reduced, enabling the prediction of target angles and ranges. These predictions serve as coarse estimates, potentially eliminating the need for DFT codebook-based beam-training. Subsequently, these estimates can be refined to obtain the candidate angle and range samples, which are critical for our NF-STAP sensing framework.} In NF-ISAC, radar functionality extends beyond conventional detection and tracking to include target classification. The integration of multimodal sensory data further enhances this classification process by providing additional contextual information, while deep learning methods automate the extraction of spatially correlated features, improving target identification and system robustness. The integration of multimodal sensory data with NF-ISAC presents a promising avenue to achieve a comprehensive and efficient fusion of sensing, communication, and artificial intelligence technologies.%In NF-ISAC, radar functionality transcends conventional roles of detection, estimation, and tracking to incorporate target classification. Feature-based methodologies, encompassing the analysis of micro-Doppler signatures, range-angle maps, and radar cross-section variations, facilitate detailed target differentiation. Furthermore, time-frequency and spectral analyses provide insights into dynamic target behaviors. Notably, deep learning techniques enable the automated extraction of spatially correlated features from NF data, contributing to robust target identification. 

%TC:ignore

\ifCLASSOPTIONcaptionsoff
  \newpage
\fi

\bibliographystyle{IEEEtran}
\bibliography{Bibliography/IEEEabrv, Bibliography/bibliography}

\renewenvironment{IEEEbiography}[1]
  {\IEEEbiographynophoto{#1}}
  {\endIEEEbiographynophoto}
  
\newpage

\begin{IEEEbiography}{Ahmed Hussain} received M.S. in Avionics Engineering from the Aerospace and Aviation Campus Kamra, Pakistan, in 2020. Currently, he is a Ph.D. student at KAUST.
\end{IEEEbiography}
\vspace*{-20\baselineskip}

\begin{IEEEbiography}{Asmaa Abdallah} received a Ph.D. in electrical engineering from the American University of Beirut, Beirut, Lebanon, in 2020. She is currently a research scientist at KAUST. 
\end{IEEEbiography}
\vspace*{-20\baselineskip}

\begin{IEEEbiography}{Abdulkadir Celik} received a Ph.D. in co-majors of electrical engineering and computer engineering from Iowa State University, Ames, IA, USA, in 2016. He is currently a senior research scientist at KAUST. 
\end{IEEEbiography}
\vspace*{-20\baselineskip}

\begin{IEEEbiography}{Ahmed M. Eltawil} received a Ph.D. degree in electrical engineering from the University of California, Los Angeles, CA, USA, in 2003. He is currently a full professor at KAUST.
\end{IEEEbiography}
%TC:endignore
\end{document}